\documentstyle[preprint,prb,aps]{revtex}
\tightenlines
\begin{document}



\draft

\title{Electron-Phonon Coupling Deduced from Phonon Line Shapes}

\author{A. Bock,\footnote{Corresponding and presenting author:
  A. Bock, Institut f\"ur Angewandte Physik und Zentrum f\"ur 
  Mikrostrukturforschung,  
  Universit\"at Hamburg, Jungiusstra{\ss}e 11, D-20355 Hamburg, 
  Germany, Tel. (040) 42838 2688, Fax (040) 42838 4368, e-mail: 
  bock@physnet.uni-hamburg.de.} 
  S. Ostertun, and K.-O. Subke}

\address{Institut f\"ur Angewandte Physik und Zentrum f\"ur 
          Mikrostrukturforschung, \\ 
          Universit\"at Hamburg, Jungiusstra{\ss}e 11, D-20355 Hamburg, 
          Germany}


\maketitle

\begin{abstract}

We investigate the Fano-type line shape of the Ba mode of  
$\rm Y_{1-x}Ca_xBa_2Cu_3O_{6+y}$ films observed in 
Raman spectra with $\rm A_{1g}$ symmetry.
The line shape is described with an extended Fano formula that allows us to 
obtain the bare phonon parameters and the self-energy effects based 
on a phenomenological description of the real and imaginary part of the 
low-energy electronic response.
It turns out that the phonon intensity originates almost entirely from 
a coupling to this electronic response with negligible contributions 
from interband (high-energy) electronic excitations.
In the normal state we obtain a measure of the electron-phonon 
coupling via the mass-enhancement factor $\lambda$.
We find $\bar{\lambda}=6.8\pm 0.5$ \% around optimum doping and only weak 
changes of the self-energy in the superconducting state.
With increased disorder at the Ba site we find a decreased intensity 
of the Ba mode 
which we can relate to a decreased electron-phonon coupling.

\end{abstract}



\narrowtext

Fano-type phonon line shapes and continua of electronic 
excitations are rather common observations in Raman experiments of
doped high-temperature 
superconductors.\cite{Cooper90,Thomsen91}
Using extended Fano models like those presented by 
Chen {\em et al.}\cite{Chen93} and 
Devereaux {\em et al.},\cite{Devereaux95} the self-energy 
contributions to the bare phonon parameters as a consequence of the 
electron-phonon interaction can in principle be obtained.
Moreover, knowledge of the self-energy effects allows one to obtain 
the mass-enhancement factor $\lambda$ as a measure of the
electron-phonon coupling strength.
This, however, requires a simultaneous description of real and imaginary 
part of the electronic response function
$\chi^e(\omega)=R^e(\omega)+i\varrho^e(\omega)$.
Such a description has recently been presented by us 
and applied to the $\rm B_{1g}$ Raman-active phonon of 
$\rm Y_{1-x}(Pr,Ca)_{x}Ba_2Cu_3O_7$
[$\rm Y_{1-x}(Pr,Ca)_{x}$-123] films.\cite{Bock99prb}
Here, we will use our description in order to investigate the Fano-type 
line shape of the Ba mode observed in $\rm A_{1g}$ Raman spectra of 
$\rm Y_{1-x}Ca_x$-123-O$\rm _{6+y}$ films near optimum doping.
This mode exhibits a pronounced asymmetry indicative of a strong 
electron-phonon coupling which is in fact obtained with our 
description.

We study epitaxial $\rm Y_{1-x}Ca_x$-123-O$\rm _{6+y}$ films 
grown by pulsed laser deposition on SrTiO$_3$(100) substrates 
in a process described elsewhere.\cite{Dieckmann96}
Standard films (x=0 ; y=1) are slightly overdoped, increased (decreased) 
doping levels are obtained by Ca substitution (oxygen reduction).
Some properties of the films are given in Table \ref{tabProp}.
Raman spectra have been taken using the Ar$^+$ laser line at 458 nm 
in a setup described elsewhere\cite{Ruebhausen97}
with the spectral resolution (HWHM) set to 3 cm$^{-1}$.
They have been corrected for the spectral response of 
spectrometer and detector and are normalized such that the $\rm B_{1g}$ 
spectrum of the film \#O4 at 18 K approaches unity above 
650 cm$^{-1}$.
$\rm A_{1g}$ spectra are obtained by subtracting the $\rm B_{2g}$ 
data from those measured in $z(x'x')\overline{z}$ geometry.
All temperatures are spot temperatures with typical heatings of 
2 K.

In order to describe the Fano-type line shape of the Ba mode 
in $\rm A_{1g}$ symmetry we 
subdivide the Raman efficiency\cite{efficiency} $I_{0}(\omega)$ into 
the electronic response and an electron-phonon interference term:
\begin{eqnarray}
    I_{0}(\omega)=\varrho_*(\omega)+\frac{C}{\gamma(\omega)
	\left[1+\epsilon^2(\omega)\right]} 
	 \left\{ \left[\frac{R_{tot}(\omega)}C\right]^2 - 
  2\epsilon(\omega)\frac{R_{tot}(\omega)}C\frac{\varrho_*(\omega)}C - 
  \left[\frac{\varrho_*(\omega)}C\right]^2 \right\} \, .
\label{eqRamInt}
\end{eqnarray}
The constant $C=A\gamma^2/g^2$ is a fit parameter for the intensity 
where $\gamma$ represents the symmetry element of the 
electron-photon vertex projected out by the incoming and outgoing 
polarization vectors and
$g$ is the lowest order expansion coefficient of the 
electron-phonon vertex describing the coupling to non-resonant intraband 
electronic excitations.
$\varrho_*(\omega)=Cg^2\varrho^e(\omega)$ and 
$R_*(\omega)=Cg^2R^e(\omega)$ are the measured electronic response 
and the real part of the electronic response function which are 
connected by Kramers-Kronig relations.
$R_{tot}(\omega)=R_*(\omega)+ R_0$ with 
$R_0=Cg(g_{pp}/\gamma)$ where $g_{pp}$ is a constant which represents 
an abbreviated ``photon-phonon'' vertex that describes
the coupling to resonant interband electronic 
excitations.\cite{Devereaux95}
The renormalized phonon parameters in the above equation are
$\gamma(\omega)=\Gamma+\varrho_*(\omega)/C$ and 
$\omega_{\nu}^2(\omega)=\omega_p^2-2\omega_p 
R_*(\omega)/C$ with
$\epsilon(\omega)=\left[\omega^2-\omega^2_{\nu}(\omega)\right]
/[2\omega_p\gamma(\omega)]$.
Note, that the interference term in Eq. (\ref{eqRamInt}) can be 
negative in contrast to the Raman efficiency.

The measured electronic response (background) 
is modeled by two contributions:\cite{Bock99prb}
$I_{\infty}\tanh (\omega/\omega_T)$ and
$I_{red}(\omega,\omega_{2\Delta},\Gamma_{2\Delta},
I_{2\Delta},I_{supp})$.
The first term models the incoherent background using a hyperbolic 
tangent and the second the redistribution below $T_c$ using 
two Lorentzians.
For the two contributions analytic expressions of the
real part of the electronic response function 
exist.\cite{Bock99prb}
In the present work, the hyperbolic tangent is cut 
off at $\omega_{cut}=8000$ cm$^{-1}$. $R_*(\omega_p)/C$ 
increases by typically 10 \% 
when the cutoff is increased to 12000 cm$^{-1}$. 
With the description according to Eq. (\ref{eqRamInt})
a measure of the electron-phonon coupling can be obtained
via the mass-enhancement factor $\lambda$, 
defined by $\lambda\cdot\omega_p = 2R_*(\omega_p)/C$.
In reference to the conventional Fano mechanism\cite{Hadjiev98b} 
the total and the bare phonon intensity are
$I_{tot}=\frac{\pi}{C} R_{tot}^2(\omega_p)$ and 
$I_{phon}=\frac{\pi}{C} R_0^2$.

To give an example, Fig. \ref{figFit} displays the results of the analysis 
of the $\rm A_{1g}$ efficiency of the film \#Ox4 taken at 18 K.
We describe this figure from top to bottom:
At the top, the measured efficiency as well as its description (solid line) 
is displayed.
For the description we used interference terms given in Eq. 
(\ref{eqRamInt}) for the Ba and the O(4) mode, Lorentzians for the Cu(2) 
and the O(2)+O(3) mode, a simple Fano formula for the $\rm B_{1g}$ 
phonon, and the two background contributions stated above.
In the second trace from the top, the phononic signal, i.e. the 
Lorentzians, the Fano profile, and the interference terms are given.
Obviously, the interference term of the Ba mode becomes negative in a 
region above $\sim 120$ cm$^{-1}$.
The electronic response $\varrho_*(\omega)$ that remains after 
subtraction of the phonons is shown below the phononic signal.
It exhibits a $2\Delta$ peak at $\sim 280$ cm$^{-1}$ as well as a 
monotonically decreasing intensity for $\omega \rightarrow 0$
and is well described with our background model 
(solid line).
At the bottom of Fig. \ref{figFit} the real part of the electronic 
response function $R_*(\omega)$ is shown.
In order to obtain $R_*(\omega)$ we have performed a numerical Hilbert 
transformation of $\varrho_*(\omega)$.
For the transformation the measured spectrum is taken as constant 
for high frequencies up to $\omega_{cut}$
and is interpolated to zero intensity at $\omega=0$; for negative 
frequencies the antisymmetry of $\varrho_*(\omega)$ has been 
used.
Evidently, the description of $R_*(\omega)$ used in the fit agrees well
with the numerically obtained data.
This is important for the determination of the self-energy effects.

At 152 K, i.e considerably above $T_c$ and below room temperature,
our description of the Ba mode in the film \#Ox4 yields
$\omega_p=121.6$ cm$^{-1}$ and $\lambda=6.3$ \%.
Similar values are obtained in the other films as given in 
Table \ref{tabProp}.
More specifically, we find a somewhat increasing bare phonon 
frequency with increasing doping with a mean value of
$\bar{\omega_p}=121.9 \pm 1.4$ cm$^{-1}$ and a mean value of the 
mass-enhancement factor of $\bar{\lambda}=6.8\pm 0.5$ \%.
The disordered film \#Ca1 clearly deviates from the others 
exhibiting a low bare phonon frequency compared with its doping value 
and an almost 50 \% smaller mass-enhancement factor.
The low frequency is in fact one indication for the presence of disorder 
at the Ba site as enlarged $c$-axis parameters are expected in this 
case.\cite{Ye94}

In order to look more closely at the peculiarities appearing in the 
disordered film we compare the temperature dependencies of the fit 
parameters of the Ba modes in the disordered film  
\#Ca1 with that of the ordered one \#Ca2 in Fig. \ref{figPhonBa}.
Beside the bare and renormalized phonon parameters also the 
self-energy contributions at $\omega=\omega_p$
are depicted.
Dashed lines are fits to anharmonic decays.\cite{Hadjiev98b} 
Except for sharpenings of the bare and the renormalized phonon 
linewidth in the ordered film \#Ca2, clear superconductivity-induced 
changes of the self-energies are not observed.
This is in good agreement with earlier results obtained on a Y-123 single 
crystal.\cite{Cooper90}
The self-energy contributions are weaker in the disordered film 
compared to the ordered one.
In particular, they differ by a factor of two above $T_c$ which is 
the same value by which the mass-enhancement factors are apart.
Noteworthy, the sharpening of the renormalized linewidth in the 
ordered film cannot be related to a suppression of the measured electronic 
response. This indicates the presence of an additional decay channel for 
the phonon.
A similar effect has been observed in case of the $\rm B_{1g}$ phonon 
in $\rm Y_{1-x}(Pr,Ca)_{x}$-123 films,\cite{Bock99prb}
where it was suspected that 
not Raman-active electronic excitations may be present in these compounds.

In the lowest panel in Fig. \ref{figPhonBa} the total intensities 
$I_{tot}$ as well as the bare phonon intensities $I_{phon}$ are given.
It turns out that the bare phonon intensities are 
negligibly small in both films at least for temperatures below 250 K.
The same finding is also observed in the other films studied.
The total intensities, on the other hand, are 30 \% stronger in the ordered 
film \#Ca2 compared to the disordered one.
This appears to be related to the stronger mass-enhancement factor 
in the ordered film which is further supported when
comparing with the factors and 
intensities of the other films given in Table \ref{tabProp}.
Regarding Table \ref{tabProp}, one finds an increasing intensity of the Ba mode 
with decreasing doping in the studied doping regime. 
This increase is carried by an increasing linewidth which 
rises from
$\Gamma=4$ cm$^{-1}$ in the film \#Ca2 up to
$\Gamma=8$ cm$^{-1}$ in the film \#Ox2.
At even lower dopings, however, the Ba mode eventually diminishes
and is no longer observed in plane-polarized Raman spectra in the 
parent compound Y-123-O$_6$.\cite{Burns91}


Isotope experiments of Y-123 have shown that the mode, 
which we have called the Ba mode so far, is indeed dominated by 
vibrations of the Ba atom with less than 20 \% admixture from the 
Cu(2) site.\cite{Strach95}
This experimental eigenvector has recently been obtained in a
linearized-augmented-plane-wave (LAPW) frozen-phonon calculation 
within a generalized gradient approximation (GGA).\cite{Kouba97}
Using the LAPW method within the local-density approximation (LDA), 
Cohen et al.\cite{Cohen90} 
find large non-local contributions 
to the electron-phonon coupling from the Ba site with 
$\lambda\leq 5.4$ \% in the Brillouin zone.
The coupling appears to be in good agreement with the
results of our extended Fano description of the ordered films.
This is somewhat surprising as the observed background 
is believed to be a consequence of the strong electronic correlations which are 
not included in LDA-type calculations.
Results of the electron-phonon coupling within the GGA are therefore 
of interest for a comparison as that method includes correlation 
effects.

To conclude, we investigate the Ba mode of
$\rm Y_{1-x}Ca_x$-123-O$\rm _{6+y}$ films observed in Raman spectra 
with $\rm A_{1g}$ symmetry.
Our Fano-type analysis reveals that this mode is entirely described by a 
coupling to low-energy electronic excitations for $T<250$ K.
The absence of this mode in antiferromagnetic Y-123-O$_6$ might 
therefore simply be a consequence of the vanishing low-energy electronic 
response.\cite{Ruebhausen97}
Our analysis yields mass-enhancement factors which appear to be 
in agreement with the result of the LAPW method within the LDA.
With increasing disorder at the Ba site, the intensity of the Ba mode 
diminshes as a consequence of a reduced coupling strength.


The authors thank D. Manske, U. Merkt, C.T. Rieck and M. R\"ubhausen 
for stimulating discussions.
S.O. acknowledges a grant of the German Science 
Foundation via the Graduiertenkolleg 
``Physik nanostrukturierter Festk\"orper''.

\begin{figure}
\caption{
  In the uppermost spectrum the Raman efficiency $I_0(\omega)$ 
  of the film \#Ox4 taken at 
  $T=18$ K in $\rm A_{1g}$ symmetry (dots)
   and the fit result (solid line) are shown.
  Below the fitted phonon profiles (solid line),
  the electronic response $\varrho_*(\omega)$ obtained after 
  subtraction of the phonons (dots), and the numerically determined 
  real part of the electronic response function $R_*(\omega)$ 
  (dots) are given.
  The analytical descriptions of the response used in the fit 
  are included as solid lines.
  The spectra are offset as indicated in brackets, all intensities are given 
  in the same units. 
  }
\label{figFit}
\end{figure}
\begin{figure}
\caption{Temperature dependence of fit parameters of the Ba mode
  mode for two Ca-doped films. Left (right) panel: disordered (ordered) 
  film. 
  Closed circles represent the bare phonon parameters 
  $\omega_p$, $\Gamma$, and $I_{phon}$, 
  open circles the renormalized values 
  $\omega_{\nu}(\omega_p)$, 
  $\gamma(\omega_p)$ and $I_{tot}$.
  Diamonds and crosses are the self-energy contributions 
  $R_*(\omega_p)/C $ and $\varrho_*(\omega_p)/ C$, respectively.
  Dashed lines are fits to anharmonic decays and dash-dotted lines 
  indicate the respective $T_c$'s of the films.
  Marker sizes represent the vertical accuracies.
  }
\label{figPhonBa}
\end{figure}
\begin{table}
\caption{Names and properties of the investigated
  $\rm Y_{1-x}Ca_x$-123-O$\rm _{6+y}$ films. 
  In contrast to the others, the film \#Ca1 exhibits 
  site-substitution disorder at the Ba site.
  The critical temperature $T_c$ is defined
  by zero resistance with $\Delta T_c\leq 2$ K.
  The hole doping per copper-oxygen plane $p$ has an accuracy of 
  $\pm0.008$ holes.
  $\omega_p$, $\lambda$, and $I_{tot}$ are the bare phonon 
  frequency, the mass-enhancement factor, and the total
  intensity of the Ba mode at 152 K, respectively.
  }
\label{tabProp}
  \begin{tabular}{lccccccc}
    
Film&x     &y&$T_c$&$p$&$\omega_p$
       &$\lambda$&$I_{tot}$   \\   
       &(\%)& & (K)    &      &(cm$^{-1}$)            
       &(\%)&(arb. units)  \\
\hline
\#Ox2 &0&$0.85\pm 0.05$&87.0&0.145&120.0&6.2&58  \\
\#Ox3 &0&$0.93\pm 0.05$&89.8&0.163&122.2&7.5&47  \\
\#Ox4 &0&$1.00\pm 0.05$&88.0&0.180&121.6&6.3&40  \\
\#Ca1 &5&$1.00\pm 0.05$&85.0&0.186&120.0&3.8&27 \\
\#Ca2 &5&$1.00\pm 0.05$&82.7&0.198&123.8&7.1&40  \\
  \end{tabular}
\end{table}  

\begin{references}
\bibitem{efficiency} The measured Raman intensity $I(\omega)$ 
is related to the imaginary part of the electronic response function 
via the fluctuation-dissipation theorem:
$ I(\omega)=\left[1+n(\omega)\right]I_{0}(\omega)
  = A\left[1+n(\omega)\right]{\rm Im}\,\chi(\omega)$,
where $n(\omega)$ is the Bose factor, $I_{0}(\omega)$ 
the Raman efficiency, 
and $A$ a proportionality constant.
\end{references}
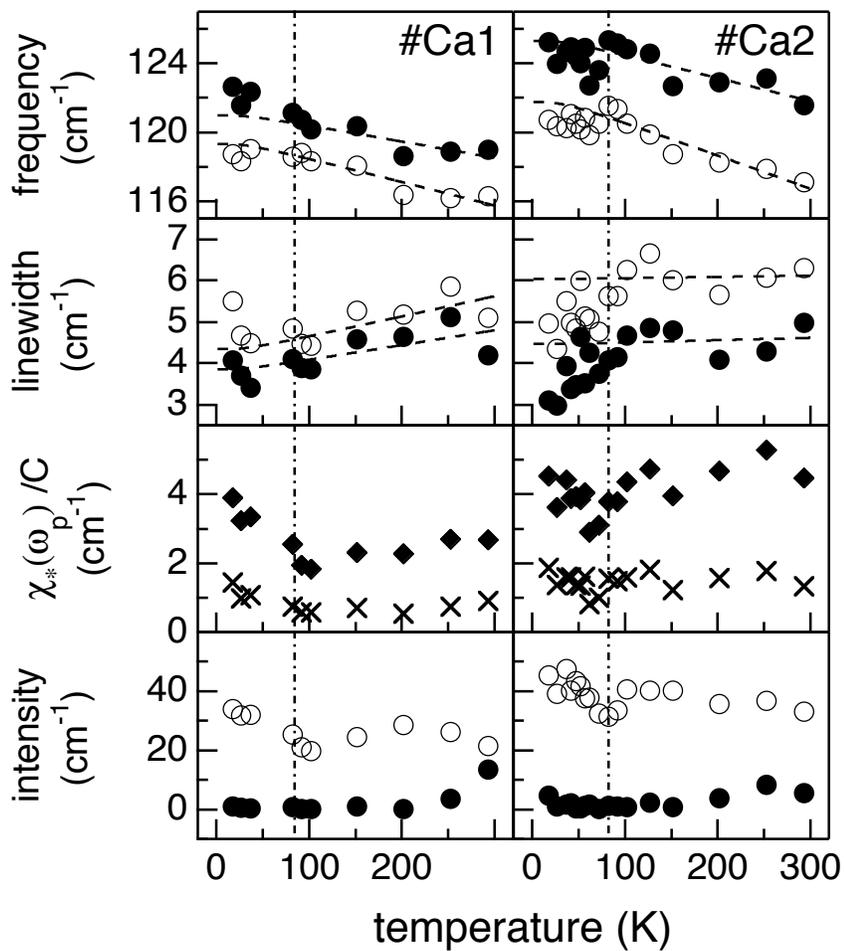
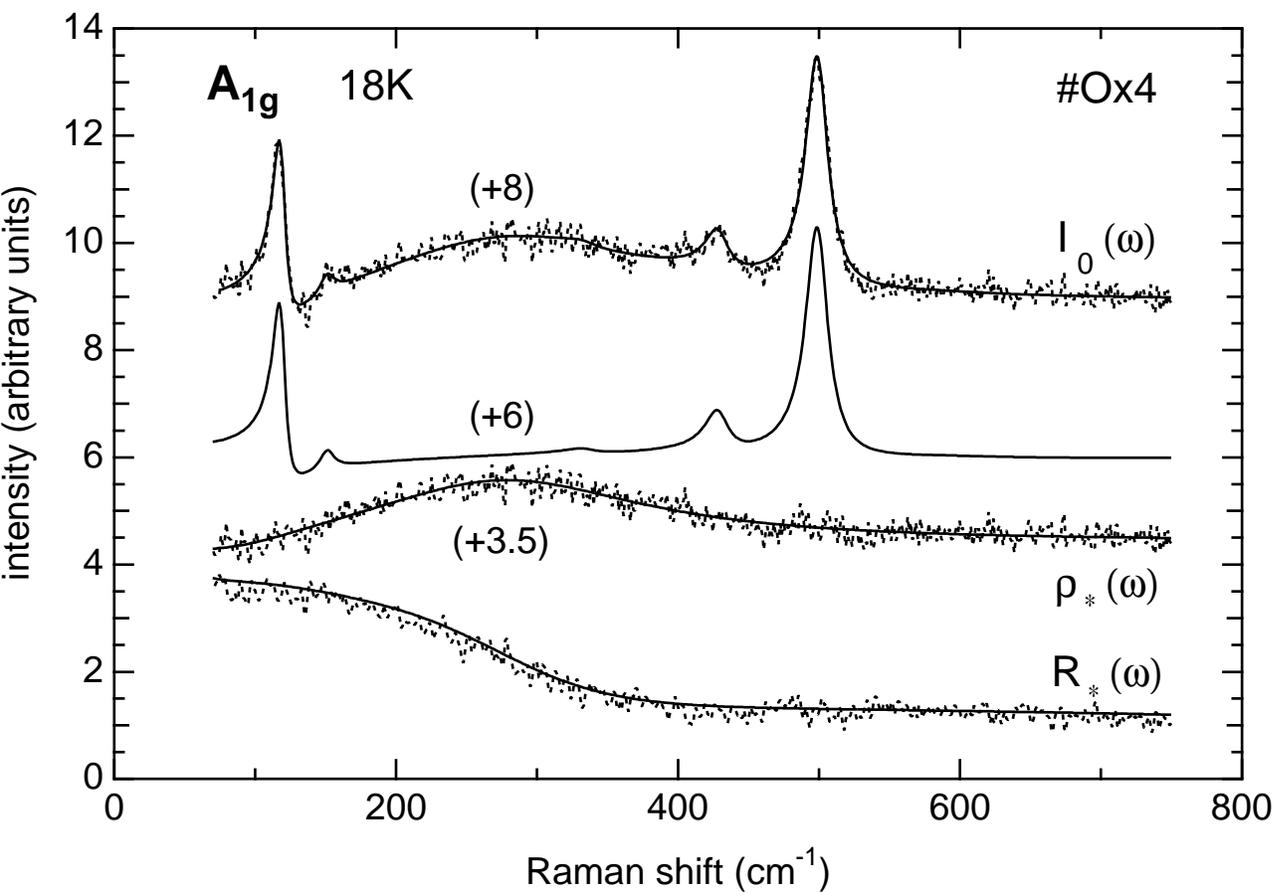
\end{document}